\documentclass[reprint,superscriptaddress,amsmath,amssymb,aps,prl]{revtex4-1}

\usepackage{times,graphicx,bbm,amsmath,amssymb}
\usepackage{epsfig,color}
\usepackage{hyperref}
\usepackage{hyperref}
\usepackage{float}
\usepackage{accents}
\usepackage{dsfont}
 \usepackage{notoccite}
\floatstyle{boxed} 

\newcommand{\bra}[1]{\langle{#1}\vert}
\newcommand{\ket}[1]{\vert{#1}\rangle}

\usepackage{color}
\usepackage{graphicx}
\usepackage{dcolumn}
\usepackage{bm}
\usepackage{float}
\usepackage{hyperref}

\begin{document}

\title{Photonic Maxwell's demon}

\author{Mihai D. Vidrighin} 
\affiliation{QOLS, Blackett Laboratory, Imperial College London, London SW7 2BW, UK}
\affiliation{Clarendon Laboratory, University of Oxford, Parks Road, Oxford OX1 3PU, United Kingdom}
\author{Oscar Dahlsten}
\affiliation{Clarendon Laboratory, University of Oxford, Parks Road, Oxford OX1 3PU, United Kingdom}
\affiliation{London Institute for Mathematical Sciences, 35a South Street, Mayfair, W1K 2XF London}
\author{Marco Barbieri} 
\affiliation{Dipartimento di Scienze, Universit\`a degli Studi Roma Tre, Via della Vasca Navale 84, 00146 Rome, Italy}
\affiliation{Clarendon Laboratory, University of Oxford, Parks Road, Oxford OX1 3PU, United Kingdom}
\author{M.S. Kim}
\affiliation{QOLS, Blackett Laboratory, Imperial College London, London SW7 2BW, UK}
\author{Vlatko Vedral} 
\affiliation{Clarendon Laboratory, University of Oxford, Parks Road, Oxford OX1 3PU, United Kingdom}
\affiliation{Centre for Quantum Technologies, National University of Singapore 3 Science Drive 2, 117543 Singapore, Republic of Singapore}
\author{Ian A. Walmsley}
\affiliation{Clarendon Laboratory, University of Oxford, Parks Road, Oxford OX1 3PU, United Kingdom}

\date{\today}

\begin{abstract}

We report an experimental realisation of Maxwell's demon in a photonic setup. We show that a measurement at the single-photon level followed by a feed-forward operation allows the extraction of work from intense thermal light into an electric circuit. The interpretation of the experiment stimulates the derivation of a new equality relating work extraction to information aquired by measurement. We derive a bound using this relation and show that it is in agreement with the experimental results. Our work puts forward photonic systems as a platform for experiments related to information in thermodynamics.

\end{abstract}

\maketitle

Maxwell's demon made its appearence in 1867 as part of a thought experiment discussing the limitations of the second law of thermodynamics~\cite{demon}. James Clerk Maxwell imagined the demon as a microscopic intelligent being, controlling a door in the wall separating two boxes that contain a gas in thermal equilibrium. The demon would use the door to filter individual particles of the gas based on their energy, producing an unbalanced particle distribution. The operation seemed to decrease the entropy of the gas without any work investment, contradicting the second law of thermodynamics. Discussions emerging from the apparent paradox played a fundamental role in revealing the relation between information and thermodynamics \cite{land, benn}. It has been shown that the amount of work that the demon can extract from the imbalance of energy between the two boxes which it can create by the sorting operation is limited by the information acquired by its measurement of the individual particles. In turn, erasure of this information from the demon's memory requires at least as much work as was extracted. Maxwell's demon has seen many reinterpretations \cite{quantDem, review} and has come to denote a system that achieves a decrease in entropy, or extraction of work by applying measurement and feedback to a system in thermal equilibrium \cite{szi, stat}. Various physical realizations of such systems have recently been demonstrated experimentally  \cite{wallDem, giappi, berut, elec}. 

Spurred by the advancement of experimental techniques, which are allowing the control of physical systems down to the single particle level, there has also been significant progress in the theoretical analysis of thermodynamics in microscopic systems, including the description of small thermal engines consisting of only a few energy levels \cite{ronnie, sandu, horo1}, fluctuation theorems \cite{jarz, saga, Sagawa, dorner, mazzola, dna, mazzolaexp}, the role of quantum coherence in thermodynamics \cite{matteo, oscar, szilard, qKos1, qKos2, quHeatMachines} and resource theories of thermodynamic transformations \cite{horo2}. As was the case in the 19th century, when the steam engine and other technologies demanded the development of thermodynamics of macroscopic systems, the modern analogues involving few or single atoms have also led to the emergence of new ideas in microscopic thermodynamics. Here we aim to explore the benefits of photonics as an experimental platform for the study of the role of information in thermodynamics. This is a powerful platform, due to the experimental tools that have been developed for photonic quantum information processing and other applications that use engineered quantum states of light.

We present an experiment analogous to Maxwell's demon, in which light plays the role of the working medium. Instead of gas particles on two sides of a wall, thermal states are prepared in two spatial light modes. We show that a measurement similar to photon subtraction \cite{grangier1, grangier2, bellini1, bellini2} on each of the optical modes and a simple conditional operation (feed-forward) can lead to a difference in average intensity between the two light modes. Following this measurement, we let the remaining light fall on two linear diodes which allow work to be extracted from the unbalanced intensities in the form of energy stored in a capacitor, a genuine, practical work reservoir. This allows us to link a microscopic measurement to macroscopic work. Interpretation of the setup as a Maxwell's demon requires a theoretical model in which the work extraction does not take place at thermal equilibrium, but as an arbitrary open system evolution. Analysing the setup leads us to derive a work-information equality, inspired by methods in \cite{saga}. This result is used to relate experimentally measurable quantities by providing a bound on the extracted work distribution in terms of measurement information. Unlike previously derived work-information inequalities, ours does not set a limit on average work extracted per cycle, but on the ratio between average work extraction and single shot fluctuations, a measure related to the strength of work discussed in \cite{sandu}. We verify experimentally that our protocol approaches the derived bound.

\begin{figure}[htpb]
\centering
\includegraphics[width=\columnwidth]{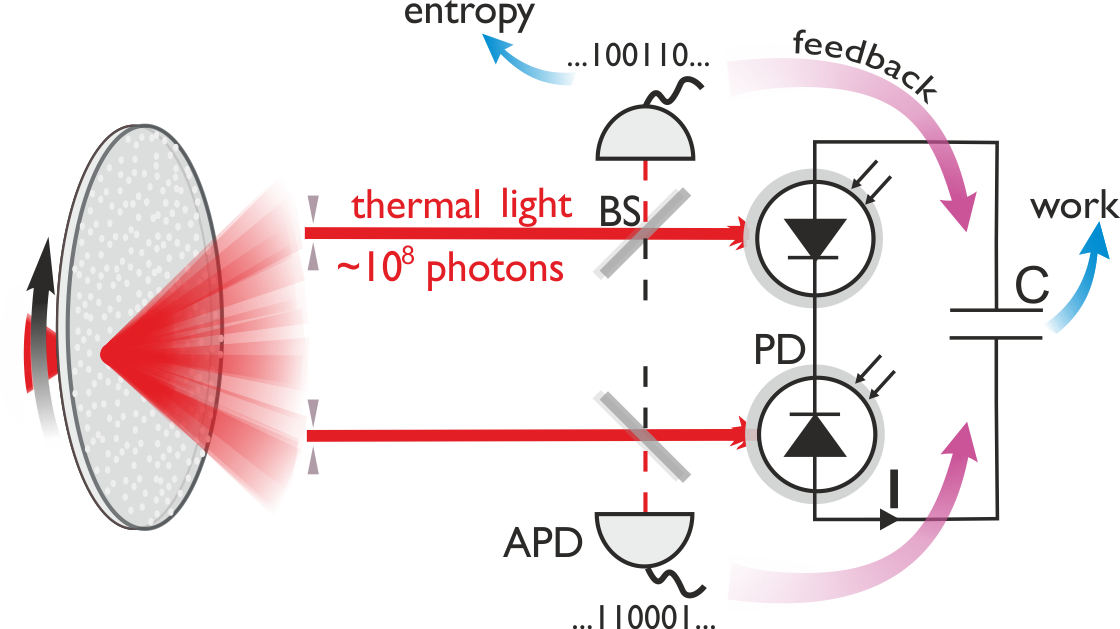}
\label{desen}
\caption{Experimental setup. An optical signal with intensity fluctuations that obey thermal statistics is obtained by collecting light from a time-varying laser speckle pattern. This is produced by focusing laser pulses onto a spinning glass diffuser (Arecchi's wheel). Two modes showing thermal statistics are selected using apertures positioned in the far field.  A measurement on these modes is implemented using beam splitters (BS) with high transmittance and single-photon sensitive click detectors (avalanche photodiodes -- APD). An electromotive source made up of linear photodiodes (PD) charges a capacitor (C). We show that a non-zero average voltage across C, required for charging a battery, can be obtained using information acquired through the APD measurement. Feed-forward can be implemented by swapping the polarity of the capacitor with an effect equivalent to swapping the two light modes after the APD measurements. In the experiment, we perform polarity swapping in post processing.}
\label{setup}
\end{figure}

\section{Setup}

Our photonics-based experiment, with schematic shown in Figure \ref{setup} follows Maxwell's demon through the characteristic steps: measurement, conditional operation and work extraction. The role of the gas in thermal equilibrium is played by two pulsed light modes, each prepared in a thermal state described by the density matrix $\left(1-e^{-\beta\, h \nu}\right)\sum_{n=0}^\infty e^{-\beta\,h \nu\, n}\ket{n}\bra{n}$ in the photon number basis where $h\nu$ is the single photon energy and $\beta$ the inverse of the thermal energy. Light pulses in the two modes have undefined phase and energy distributed according to the Boltzmann distribution. We prepare these states by collecting light from a variable laser speckle pattern, which is produced using a spinning glass diffuser as depicted in Figure \ref{setup}. This type of source is known to produce light with thermal fluctuations \cite{arecchi,shih,bellini3,parigi} offering the possibility to obtain much higher intensities that those achievable by selecting the emission into a single mode of a thermal lamp.

The demon's measurement is similar to photon subtraction  \cite{grangier1, grangier2, bellini1, bellini2}. Each thermal light mode propagates through a high transmittance beam splitter and the reflected light is coupled to a single-photon detector. The state inferred form observing a detection event has a different average number of photons than the incident light mode, increased or decreased depending on whether the incident light has super-poissonian or sub-poissonian statistics. When this measurement is applied to single-mode thermal light, if the probability of the photon detection is made very small, the average number of photons is doubled in the cases when a photon detection happened \cite{bellini2}. Consider the extreme case in which an incident light pulse is either vacuum or populated by a large number of photons. Detection of even a single photon from this incident pulse would distinguish with certainty the two possibilities. In a similar manner, our measurement determines, although with remaining uncertainty, the energy fluctuations of a thermal light mode. And if the detection outcomes are ignored, the effect of the measurement amounts to a negligibly small loss introduced by the high transmittance beam splitter. In our setup we use on/off detectors which distinguish between vacuum and one or more photons, giving a simple binary intensity measurement. We do not restrict the photon detection rate (the number of times that the demon's detector fires per pulse) to small values as is commonly the case in quantum optics experiments. Thus the energy fluctuation resolving power of our measurement can be tuned by fixing the amount of light sent towards the single-photon detectors, which sets the photon detection rate (see Appendix A). Feed-forward can be implemented by swapping the two thermal light modes conditioned on the output of the demon's measurement, so that on average there is more energy on one side. In this way an asymmetric energy distribution can be created from two equally populated thermal light modes.

For work extraction we propose the detection of the two thermal light modes on two photodiodes, connected with opposing polarities such that on average they produce zero voltage. This photodiode circuit includes a capacitor which is charged according to the fluctuating energy difference between the two detected modes. If an unbalance in the energy distribution of the two modes is produced by measurement and feed-forward, the capacitor will have a non-zero average charge which can be used to charge a battery, extracting single cycle work. We choose this setup for its conceptual simplicity, not aiming to realise an optimal work extraction strategy.

In practice, we make some simplifications to the measurement and controlled operations, aiming to provide a proof of principle implementation. Firstly we note that photon subtraction can be implemented with imperfect photon detectors and in Appendix A we show that the effect of the demon's measurement does not depend on the detection efficiency. Therefore, we are not required to implement a beam splitter with reflectance on the order of the inverse average photon number in the thermal light modes. Instead, we implement a reflectance of $5\cdot10^{-3}$, small enough for the average effect on the transmitted light to be negligible compared to the thermal fluctuations. The detection rates of the demon's measurement are then regulated using variable absorbers. Secondly, swapping of the thermal light modes can be implemented on-line using a variable beam splitter triggered by the single-photon detectors. However, as long as the two thermal light modes are well balanced, the symmetry of the setup is such that the swapping of the two light modes is indistinguishable from switching the polarity of the capacitor. Thus, we replace feed-forward by a logical operation on the experimental data, switching the sign of the measured capacitor voltage as a function of the single-photon detector outputs.

\section{A non-equilibrium work-information equality inspired by the experiment}

Non-equilibrium work relations such as the celebrated Jarzynski equality \cite{jarz} link non-equilibrium processes to equilibrium quantities like the thermal free energy. One equality derived by Sagawa and Ueda incorporates the effect of measurement and conditional operations \cite{saga, Sagawa}, providing a way to derive work-information bounds in Maxwell's demon type scenarios. The effectiveness of a demon's measurement is included in this expression through the mutual information quantifying correlations between measurement outcomes and the measured system. When the initial energy state of the measured system is $i$ and the measured outcome is $m$ the point-wise mutual information is $I=\log\left(p(m\vert i))-\log(p(m)\right)$. Here, by $p(m)$ we denote the probability of outcome $m$ and by $p(m\vert i)$ the conditional probability of $m$ given $i$. The theorem by Sagawa and Ueda reads $\langle e^{\beta (W-\Delta F)-I}\rangle=1$ where $W$ is single cycle work extraction and $\Delta F$ is the free energy difference between the final and initial states of the working system. Jensen's inequality can be used to derive from this a bound on extracted work: $\beta W \leq \beta \Delta F+I$ showing that information extracted by measurement allows for work extraction, even without free energy consumption. We note that the entropy of the measurement register, which can be readily estimated from a set of measurement outcomes, provides an upper bound to the mutual information $I$. The theorem by Sagawa and Ueda holds for work extraction scenarios where local detailed balance applies. 

Given the physical complexity of the work extraction setup presented here, this model does not apply to our experiment. To find a similar relation between information gain and extracted work we require a theoretical model allowing full generality of the work extraction operation. We seek inspiration in our experiment, noting that work extraction can only be performed by acting efficient feed-forward. This observation is with respect to the situation when the demon's measurement is simply ignored, when no work is extracted. We aim to include this scenario in our theoretical model. This leads us to the following equality, the first main theoretical result of our work:
\begin{equation}
\langle e^{\beta W-I}\rangle_{\text{f}} = \langle e^{\beta W}\rangle_0
\label{main}
\end{equation}
Here, the left hand term is an average (denoted 'f') corresponding to the situation with feed-forward, controlled by the output of the measurement whose efficiency is quantified by the mutual information $I$. The right hand side (denoted '0') is an average corresponding to the same system, but where the measurement and feed-forward steps are missing. This means that the measurement outcomes are simply ignored when the measurement is on average non-disturbing, such as in our setup (the effect of the high transmittance beam splitter is negligible). As we show in the following section, Equation \ref{main} can lead to useful results. The model and assumptions under which this equality is derived are detailed in the Proofs section. We note that this is valid when the feedback is an energy conserving unitary operation acting only on the thermal system measured by the demon and when a non-disturbance condition applies to the measurement, which we show is the case for our setup.

\section{A bound relating work extraction to measurement information}

We now use Equation \ref{main} to derive a bound applicable to measurable quantities in the experiment. Let $U$ denote the voltage created across the capacitor $C$. The external work reservoir, a battery, can be charged by connecting it to the capacitor. The energy transfer from capacitor to battery depends on the voltage of the battery $U_0$, which must be different from zero for work to be extracted. This energy transfer is $W=C(U-U_0)U_0$ where $C(U-U_0)$ is the charge transported across the battery against the potential difference $U_0$ after the capacitor and battery are connected. Inserting this into Equation \ref{main} we get $\langle e^{\beta C U_0 U-I}\rangle_{\text{f}} = \langle e^{\beta C U_0 U}\rangle_0$ and using Jensen's inequality, $\beta C U_0\langle U\rangle_{\text{f}} -\langle I\rangle \leq \log (\langle e^{\beta C U_0 U}\rangle_0)$.  We define $\epsilon=\beta C U_0$. Since the inequality is valid for any value of $U_0$, we can find the tightest bound on $\langle U \rangle_{\text{f}}$ by optimizing with respect to $\epsilon$. While this can be done for any distribution of $U$, a particularly relevant case is that when, ignoring the demon's measurement outcomes, $U$ is normally distributed with mean $\langle U\rangle_0$ and standard deviation $\sigma(U)_0$. Then $\log(\langle e^{\epsilon U}\rangle_0) = \epsilon \langle U \rangle_0 + \frac{\epsilon^2 \sigma(U)_0^2}{2}$ and optimizing over $\epsilon$ yields the bound $\frac{\vert\langle U \rangle_{f}-\langle U \rangle_{0}\vert}{\sigma(U)_0} < \sqrt{2 \langle I\rangle}$. Using the relation between work extracted and voltage on the capacitor, we can rewrite this equation in terms of work: 

\begin{equation}
\frac{\vert\langle W \rangle_{f}-\langle W \rangle_{0}\vert}{\sigma(W)_0} < \sqrt{2 \langle I\rangle}.
\label{bound}
\end{equation}

A special feature of this bound is that it does not contain $\beta$ as a scaling factor, with work fluctuations defining the scale instead. The fact that the mutual information appears inside a square root is not surprising since as we repeat the same protocol, average extracted work and mutual information scale linearly with the number of repetitions while the standard deviation of the work distribution scales with the square root of the number of repetitions. Since for many repetitions of the same protocol the total work distribution will tend to normality by the law of large numbers, we can use the bound given by Equation \ref{bound} for any well behaved single shot work distribution.

\section{Experimental results}

We first measure the distribution of pulse energy in the thermal light modes, by sampling 4000 consecutive pulses. The creation of thermal states is certified by estimating the intensity autocorrelation of the light pulses at zero delay, $g^{(2)}(0)$ \cite{multimode}. For an ideal single mode thermal state this should yield $g^{(2)}(0)=2$. Using the measured pulse energy values, we repeatably obtained $1.9<g^{(2)}(0)<2$ and no cross correlation in the pulse energy of the two modes. The intensity of the two optical modes was balanced: the deviation from zero of the average voltage produced by the photodiode source in the absence of feed-forward was less than $0.3\%$ of the standard deviation of this voltage, corresponding to thermal fluctuations.

\begin{figure}[h]
\centering
\includegraphics[width=0.5\textwidth]{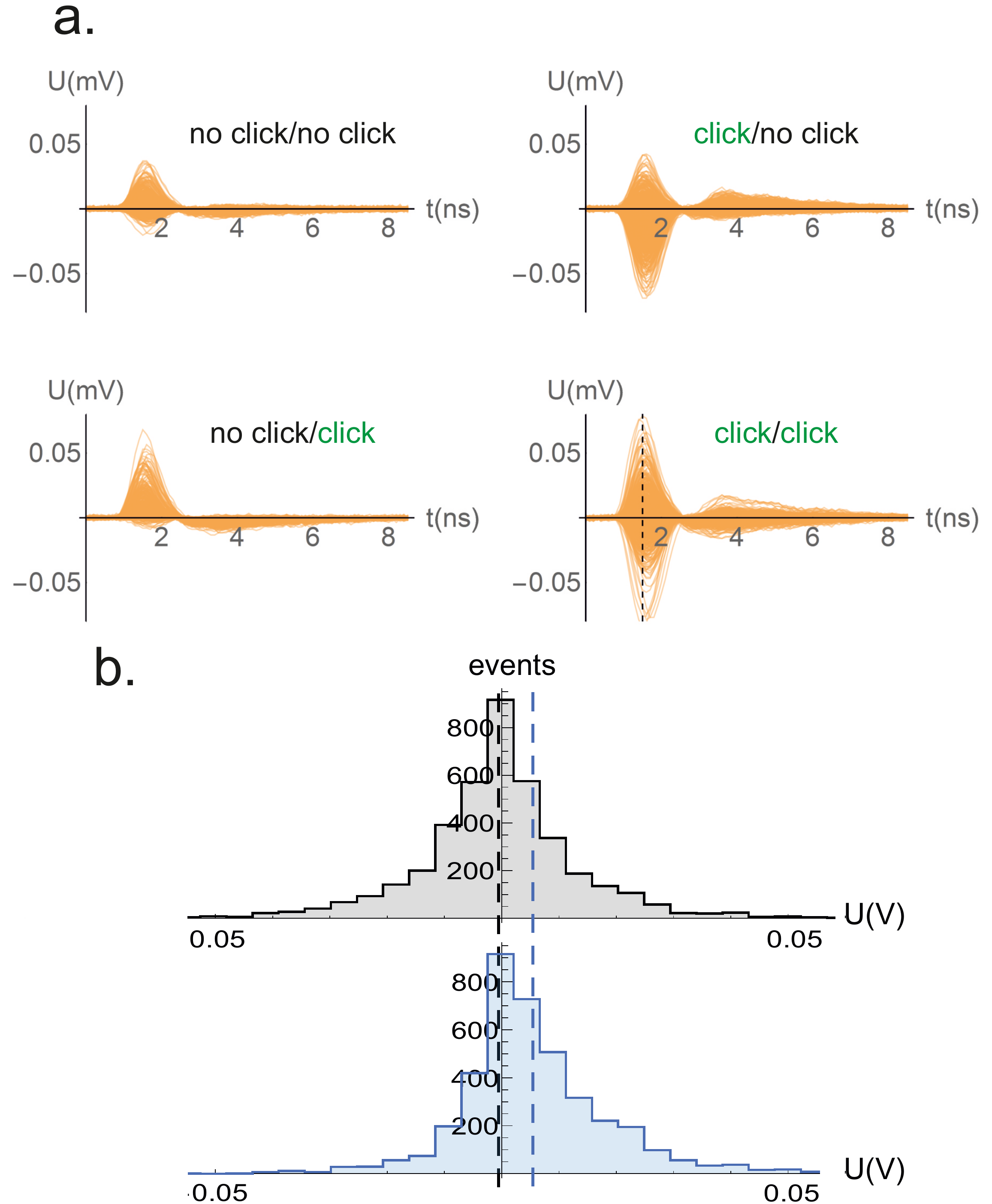}
\caption{Measured voltage on capacitor $C$. Oscilloscope traces showing the voltage on C created by the linear photodiode electromotive source. The traces are filtered by measurement outcomes. (a) 4000 traces, sorted according to binary signals from the two APDs implementing the demon's measurement (click, corresponding to photon detection or no click, corresponding to vacuum). The black dashed line indicates the time at which the maximum voltage was sampled. (b) Histogram depicting the distribution of the maximum voltage on C. Gray -- the APD outputs are ignored; blue -- a logical operation conditioned by APD outputs is implemented: the sign of the trace is flipped when the two APD signals are click and no click, respectively. The dashed vertical lines are showing the averages of the two distributions. There is a clear displacement in the average voltage when a conditional operation is applied, showing that feed-forward can produce a non-zero average voltage on the capacitor. The firing rates for the two click detectors were $p_1=0.702\pm0.008$ and $p_2=0.311\pm0.008$ respectively.}
\label{exp}
\end{figure}

In our experiment, we record oscilloscope traces of the voltage across capacitor $C$, at the same time recording outcomes of the APD detection, as depicted in Figure \ref{setup}. In Figure \ref{exp} we illustrate how the voltage depends on the APD signals and how, while the average voltage is close to zero when we ignore the demon's measurement, it becomes significantly different from zero (relative to fluctuations) when we apply a sign flip conditioned on the APD measurement outcomes. This conditional operation emulates unitary feedback on the two thermal optical modes.

The demon's measurement can be tuned by varying the amount of light sent towards the single-photon detectors. This allows us to change the photon detection rates from zero to one (detections per pulse), and as we show in Appendix A, the imbalance that can be created in the two optical modes depends only on these rates. One might expect that the imbalance that can be obtained between the two modes be highest when the detection probabilities are around $1/2$, which corresponds to the highest entropy (information content) of the measurement register. However, we show in Appendix A that the probabilities that maximise the average unbalance are actually $1/3$ and $2/3$ for the two arms respectively. For our experiment, we set one of the arms to a detection rate of $0.311 \pm 0.008$ and scan the rate $p_1$ corresponding to the other arm. There are two different ways in which feed-forward could be applied, corresponding to a change of the sign of the voltage across the capacitor corresponding to either one of the two asymmetric outputs of the demon's measurement. Which of these two strategies is optimal depends on  the choice of $p_1$.

\begin{figure}[htpb]
\center
\includegraphics[width=0.5\textwidth]{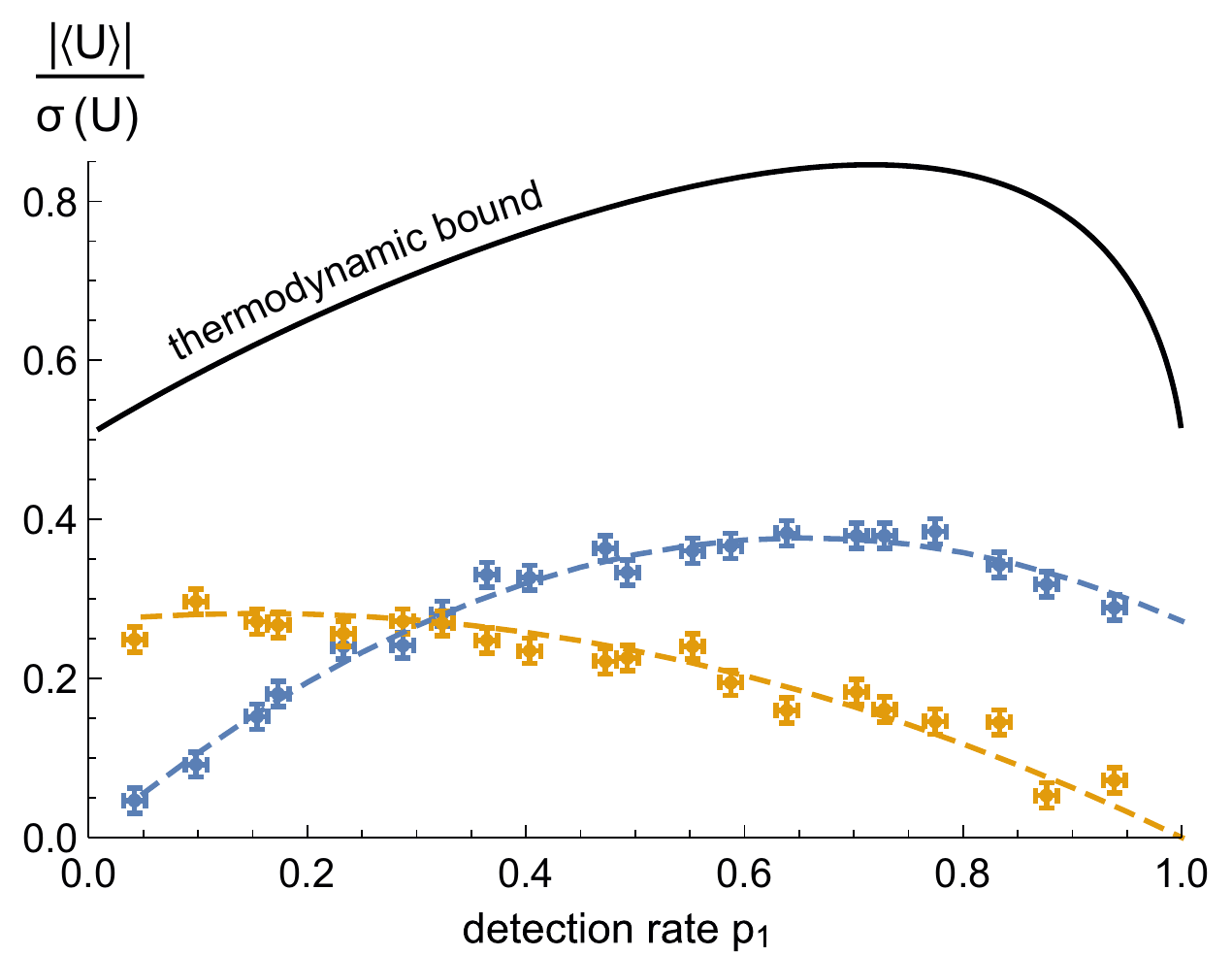}
\caption{Extracted work linked to information. Experimental points showing the absolute average voltage $U$ produced on capacitor C, which is directly proportional to the work that can be extracted by discharging the capacitor into a battery, weighted by the standard deviation of this voltage $\sigma (U)$, as a function of measurement settings. The demon's measurement is tuned by changing the APD detection rates. We choose a detection rate of $0.311 \pm 0.008$ for the second APD and tune the rate $p_1$ of the first detector between zero (no clicks) and one (a click for every pulse). This setting was chosen because the maximum average voltage is obtained when the two detectors have probabilities of firing $2/3$ and $1/3$, respectively. Blue and orange points correspond to two types of feedback: flipping the voltage when the measurement yields click/no click (blue) and flipping the voltage when the measurement yields no click/click (orange). The dashed lines are simple models based on the average number difference in two multimode thermal states  corresponding to a second order autocorrelation function $g^{(2)}(0)=1.9$, as  measured in the experiment. Error bars are estimated by binning the experimental data and computing the variance of the values shown in the figure. The black line gives the bound established in Equation \ref{bound} in terms of mutual information, the computation of which is detailed in Appendix B.}
\label{exp2}
\end{figure}

In Figure \ref{exp2} we show the average voltage that can be obtained with different measurement and feedback settings and compare this to the thermodynamic bound introduced in Equation \ref{bound}. If no information were acquired by the demon's measurement, the bound would demand that no work should be extracted. We thus find that the work extraction is caused by the acquisition of information in our setup and that the strategy that we apply for using the information yields results that are close to a thermodynamic limit. 

\section{Discussion}

The experiment presented in this work links very different regimes: the single photon regime, a measurement yielding single bit outcomes, intense light fields with thermal occupation number corresponding to very high temperatures and the room temperature system composed of detectors, electronics and the environment. We are able to show that the setup is like Maxwell's demon, in the sense that work extraction is limited by information acquired through measurement. However, the quantity bounded by information is not the average work extracted per cycle, as is usually the case \cite{land, benn, review, saga} but the ratio between the average work extraction and the standard deviation of the single cycle work distribution. Rather than defining the energy efficiency of the demon's control, this quantity describes the purity of the work produced, being related to the concept of strength of work introduced in \cite{sandu}. This quantity is of relevance in scenarios in which low fluctuations are important, such as cooling experiments. 

Our work demonstrates how photonics can provide a valid experimental platform for thermodynamic scenarios. This offers the perspective of moving into the quantum domain, to further explore the interface between quantum information and thermodynamics thanks to the capability to engineer the wave-function of multi-photon states. In addition, the techniques presented can be extended to opto-mechanical oscillators \cite{optoMech} and spin-ensembles \cite{atoms}, where single-particle operations can be used to study the link between information and thermodynamics in stationary matter systems.

\section{Experimental methods}

Optical detection using standard silicon linear photodiodes requires that our source of thermal light have a high average photon number per pulse: the source that we used, described in Figure \ref{setup} yields a number of photons per pulse on the order of $10^8$. We achieved this by scattering 4mJ pulses from an amplified Ti:Sapphire laser to produce a laser speckle pattern and multimode optical fibers with core size of $25 \mu\text{m}$ to collect the light. A laser speckle pattern with correlation area larger than the collection aperture is required in order to observe single mode thermal statistics. A fine grit glass diffuser and a relatively tight focusing of the laser, using a $5\,\text{cm}$ lens yielded an appropriate speckle pattern. The fiber apertures were positioned $15\,\text{cm}$  away from the glass diffuser in the speckle pattern. We strongly chirped the laser pulses in order to avoid nonlinear damage of the glass diffuser which tends to smooth the diffusing surface. The intensity autocorrelation $g^{(2)}(0)$ of the produced light was repeatedly measured to be $1.9<g^{(2)}(0)<2$. The $g^{(2)}(0)$ is smaller than $2$ for multimode thermal light. The mode number in this case is given by $1/(g^{(2)}(0)-1)$ \cite{multimode}. Multimode states are discussed in Appendix A.

We used a capacitor with capacitance $C=2\text{pF}$ and measured the voltage created across it by each laser pulse using an oscilloscope. The transmittance of the measurement beam splitters was $T=99.5\%$. We lowered the reflected power using variable neutral density absorbers before coupling the signal into fibers leading to the APDs, allowing the tuning of the APD photon detection rate between zero and one detections per pulse. As we show in Appendix A, for high values of $T$, uniform losses in the reflected light are equivalent to a lower reflectance, in terms of the effect that the measurement has on the measured light. In terms of overall efficiency, the $0.5\%$ loss due to the beam splitter is a negligible effect, making our setup operationally equivalent to what we would obtain with a much higher transmittance.

\section{Proofs}


To define work extraction in our model of the experimental setup, we divide the model system into three parts. The first part starts in thermal equilibrium, with inverse temperature $\beta$, having no correlations with the rest of the system. It corresponds to the two thermal light modes. The second part is the battery, or work reservoir and the third part is defined as everything that is neither the battery nor the thermal system, which corresponds in our setup to the photodiodes, capacitor and environment. We define work extracted as the energy increase of the battery. Measurement and feed-forward operate only on the first part, the thermal system (two light modes). Feed-forward is described by a unitary operation with no energy cost, conditioned on the measurement outcome. This is an appropriate description of a mode swapping operation, which can be in principle implemented by a variable-reflectivity beam splitter that switches without energy consumption. Finally, we impose a condition on the measurement, as is detailed below. This is related to the notion of non-disturbance and applies to the measurement implemented in our experiment. 

Let us denote the initial energy eigenstate of the thermal system (on which measurement and feedback are performed) $\ket{i}\bra{i}$, with energy $E_i$. For the work extraction system (second and third parts as described above), we use $\ket{j}\bra{j}$ to denote the initial state and $E_{j}$ to denote the corresponding energy, excluding the energy of the work reservoir. For the final state of the \textit{whole} system we use $\ket{f}\bra{f}$ to denote the state and denote $E_f$ the corresponding energy, again excluding the work reservoir. The extracted work is defined $W=E_i+E_j-E_f$. The initial probability distribution of the system's state in the basis defined above is $p(i, j)=\frac{1}{Z}e^{-\beta E_i}p(j)$. Let the demon's measurement outcome be denoted $m$ and the effect of the measurement be defined by the non-linear map $\mathcal{M}_m$ yielding normalized states:  $\mathcal{M}_m(\ket{i}\bra{i})=\sum_k M_k^{(m)}\ket{i}\bra{i} M_k^{(m)\dagger}/p(m|i)$ where $M_k^{(m)}$ are the measurement operators corresponding to output $m$. These are positive operators normalized such that $\sum_{m,k} M_k^{(m)\dagger}M_{k}^{(m)}=\mathds{1}$. The feedback is represented by unitary operators $U_m$. The work extraction that takes place after the feedback is modeled as an energy conserving evolution of the whole system, according to a unitary operator $V$.  Pointwise mutual information is defined $I=\log(p(m\vert i))-\log(p(m))$. We denote $x\equiv\{f,m,i,i^\prime\}$. Using these definitions, the left hand side of Equation \ref{main} is 
\begin{equation}
\begin{split}
\sum_x\, &p(f,m,i,j)\,e^{W-I}=\\
\sum_x\, &p(f\vert m, i, j) p(m\vert i) p(i, j)\,e^{\beta W-\log(p(m\vert i))+\log(p(m)}=\\
\sum_x\, &p(f\vert m, i, j) p(m)\frac{1}{Z}\,e^{-\beta E_i}p(j)\,e^{\beta(E_i+E_j-E_f)}=\\
\sum_x\, &p(f\vert m, i, j) p(m) p(j)\frac{1}{Z}e^{\beta E_j-\beta E_f}
\end{split}
\label{split}
\end{equation}
where in the second line we used Bayes' rule. We can write out the probability $p(f\vert m, i, j)=\bra{f} V \left( U_m \mathcal{M}_m (\ket{i}\bra{i}) U_m^\dagger\!\otimes\!\ket{j}\bra{j}\right) V^\dagger \ket{f}$. As we show below, for the measurement implemented in our setup, we have
\begin{equation}
\sum_i\, \mathcal{M}_m(\ket{i}\bra{i})=\mathds{1}.
\label{condition}
\end{equation}
We note that this condition holds for any non-disturbing measurement, when $\mathcal{M}_m(\ket{i}\bra{i})=\ket{i}\bra{i}$. Using this and $U_m U_m^\dagger=\mathds{1}$ and $\sum_m p(m)=1$ we get the following expression for the left hand term of Equation \ref{main}:
\begin{equation}
\begin{split}
&\langle e^{\beta W-I}\rangle_{\text{f}}=\\
&=\frac{1}{Z}\sum_m\, p(m) \sum_{j,f} \bra{f}V\left(\mathds{1}\otimes p(j) \ket{j}\bra{j}\right)V^\dagger \ket{f}e^{E_j-E_f}\\
&=\frac{1}{Z}\sum_{j,f} \bra{f}V\left(\mathds{1}\otimes p(j) \ket{j}\bra{j}\right)V^\dagger \ket{f}e^{E_j-E_f}.
\end{split}
\end{equation}
Thus using the assumptions of our model, we have obtained an expression independent of the details of the measurement and feedback operations. Similarly, the right hand side of equation \ref{main} is
\begin{equation}
\begin{split}
&\langle e^{\beta W}\rangle_{\text{0}}=\sum_{i,j,f}\, p_0(i,j,f)\,e^{\beta W}=\\
&\sum_{i,j,f}\, \bra{f} V\left(\frac{1}{Z}e^{-\beta E_i}\ket{i}\bra{i} \otimes\!p(j)\ket{j}\bra{j}\right)V^\dagger \ket{f} e^{\beta(E_i+E_{j}-E_f)}=\\
&\frac{1}{Z}\sum_{j,f} \bra{f}V\left(\mathds{1}\otimes p(j) \ket{j}\bra{j}\right)V^\dagger \ket{f}e^{E_j-E_f}.
\end{split}
\end{equation}
This is the same expression as for the left hand term. Therefore Equation \ref{main} holds for our model.

We now show that Equation \ref{condition} holds for our measurement setup, described in the main text and depicted schematically in Figure \ref{setup}. Here we derive this for the idealized, lossless measurement and in Appendix A we show that the experimental implementation is equivalent to this model. For our setup, the state with energy $E_i$, according to the notation described above, is the state with $i$ photons.\\
Let us start with the measurement outcome corresponding to no photons detected ($m=0$). When this outcome is recorded given the input state $\ket{i}\bra{i}$ the input is left unchanged: if no photons were detected in the reflected arm of the beam splitter, all photons must have been transmitted. Therefore we have $\sum_i\, \mathcal{M}_0(\ket{i}\bra{i})=\mathds{1}$. 
In the case of the outcome corresponding to a photon detection ($m=1$), the situation is not as simple. The measurement operators describing the detection of k photons are, in the photon number basis, $M_k=\sum_i \sqrt{{i \choose k} T^{i-k}(1-T)^k} \ket{i-k}\bra{i}$ with $T$ the beam splitter transmittance. The probability of outcome $m=1$ when the initial state is $\ket{i}\bra{i}$ is $1-T^i$ (with probability $T^i$ all photons are transmitted and the measurement outcome is $m=0$). We thus have
\begin{equation}
\begin{split}
\sum_i\, &\mathcal{M}_1(\ket{i}\bra{i})=\sum_{i,k\geq 1} M_k \ket{i}\bra{i} M_k^\dagger/(1-T^i)\\
=&\sum_{i,k\geq 1}{i \choose k}\frac{T^{i-k}(1-T)^k}{1-T^i}\ket{i-k}\bra{i-k}\\
=&\sum_{i,k\geq 1}{i+k \choose k}\frac{T^i(1-T)^k}{1-T^{i+k}}\ket{i}\bra{i}.
\end{split}
\end{equation}
We can bound the coefficients of $\ket{i}\bra{i}$ from above using 
\begin{equation}
\begin{split}
\frac{1}{1-T^{i+k}}&\leq\frac{1}{1-T^{i+1}}\Rightarrow\\
\sum_{k\geq 1}{i+k \choose k}\frac{T^i (1-T)^k}{1-T^{i+k}}&\leq\sum_{k\geq 1}{i+k \choose k}\frac{T^i (1-T)^k}{1-T^{i+1}}=1/T
\end{split}
\end{equation}
and from below, using the first term of the sum (all terms are positive):
\begin{equation}
\sum_{k\geq 1}{i+k \choose k}\frac{T^i (1-T)^k}{1-T^{i+k}}>{i+1 \choose 1}\frac{T^i (1-T)}{1-T^{i+1}}.
\end{equation}
As $T\rightarrow 1$, the limit of both the upper and the lower bound is $1$. This shows that $\lim_{T\rightarrow 1}\sum_i\,\mathcal{M}_1(\ket{i}\bra{i})=\mathds{1}$, so for high transmittance beam splitters, the condition given by Equation \ref{condition} applies approximatively up to an error of the order $|1-T|$.

\section{Acknowledgments}

This work was supported by the UK Engineering and Physical Sciences Research Council (EPSRC EP/K034480/1). M.V. is supported by the Controlled Quantum Dynamics DTC. O.D. is supported by EPSRC, the John Templeton Foundation, the Leverhulme Trust, the Oxford Martin School, the NRF (Singapore), the MoE (Singapore) and EU Collaborative Project TherMiQ (Grant Agreement 618074). M.B. is supported by a Rita Levi-Montalcini fellowship of MIUR. M.S.K. is supported by EPSRC and the Royal Society. We thank B. Smith, A. Gardener, D. Jennings, M. Mitchison, M. Vanner, S. Kolthammer and R. Tetean for useful discussions.

\section{Appendix A}
 
In this appendix we calculate the effect that the single photon level measurement described in Figure \ref{setup} has on a mode in a thermal state. These calculations are used for the theoretical predictions depicted in Figure \ref{exp2}.
 
Let $T$ be the measurement beam splitter's transmittance, $R=1-T$ the beam splitter's reflectance and $\eta$ the efficiency of the single photon detection. A single mode thermal state can be written $(1-\lambda)\sum_i \lambda^i \ket{i}\bra{i}$ in the photon number basis. The POVM element corresponding to detection of k photons in our setup is $M_k=\sum_i \sqrt{{i \choose k} T^{i-k}(R \eta)^k} \ket{i-k}\bra{i}$ with $k=0$ corresponding to no detected photons and $k\geq 1$ corresponding to at least one photon being detected. Applying these operators to the initial state, we obtain that the probability for any of the $k\geq 1$ outcomes to occur is $p_{k\geq 1}=\frac{\lambda R \eta}{1-(1-R\eta)\lambda}$ and the mean number of photons in the corresponding post measurement state is $\langle n\rangle_{k\geq 1}=2 T\frac{1-(1-R\eta/2)\lambda}{1-(1-R\eta)z\lambda}\langle n\rangle_t$ where $\langle n\rangle_t$ is the average photon number in the initial thermal state. We thus have that $\frac{\langle n\rangle_{k\geq 1}}{\langle n\rangle_t}=(2-p_{k \geq 1})T$. The number of photons in the state created when outcome $k=0$ is observed is $\langle n\rangle_{k=0}=(1-p_{k\geq 1})\langle n\rangle_t T$. For high transmittance beam splitters, $T$ can be approximated with $1$. We thus see that the average number of photons is a simple function of the detection probabilities and that these probabilities can be tuned by changing the efficiency $\eta$, given high $T$.
 
We can now calculate the expected average photon number difference when the two light modes are swapped according to the demon's measurement outcomes. We denote $p_1$ and $p_2$ the probabilities for photon detection to occur in the two modes respectively. The average difference in the two arms has the following contributions: for no detector firing, the difference is $(p_2-p_1)\langle n\rangle_0$ with probability $(1-p_1)(1-p_2)$; for only the first detector firing, $(p_2-p_1+1)\langle n\rangle_0$ with probability $p_1(1-p_2)$; for only the second detector firing, $(p_2-p_1-1)\langle n\rangle_0$ with probability $(1-p_1)p_2$ and for both firing, $(p_2-p_1)\langle n\rangle_0$ with probability $p_1 p_2$. The maximum average photon number difference is obtained if the modes are switched in the third of these cases, yielding $\langle n\rangle_1-\langle n\rangle_2=16/27\langle n\rangle_0$ for detection probabilities $p_1=1/3$ and $p_2=2/3$.
 
The statistics of the light produced in our setup is slightly multi-mode, which is indicated by the second order autocorrelation function measured, $1.9<g^{(2)}(0)<2$. When the thermal light is not single mode, the average number of photons prepared by measurement is different from the single mode case treated above. Let the input state be a mixture of k thermal modes. We denote the probability for the single photon detector to fire if only one of the $k$ thermal modes was probed $q$ and the probability for the detector to fire when all modes are probed $p$. The probability for the detector not to fire when all k modes are measured is $1-p=(1-q)^k$. Using the result above, we know that the average number of photons created by this outcome is $(1-q)\langle n\rangle_t = (1-p)^\frac{1}{k}\langle n\rangle_t$. Using energy conservation, we find the average number of photons corresponding to a photon detection outcome: $\frac{1-(1-p)^{\frac{k+1}{k}}}{p}$.
 
\section{Appendix B}
 
Here we calculate the average mutual information characterizing the measurement described in Figure \ref{setup}. We show that for high transmittance $T\approx 1$ of the measurement beam splitter and high average photon population of the initial states, this quantity depends only on the single photon detection probabilities. The calculation presented here is used to depict in Figure \ref{exp2} the bound given by Equation \ref{bound}.
 
The average mutual information is
\begin{equation}
I=\sum_{m,i} p(m,i) \left(\log(p(m\vert i))-\log(p(m))\right)
\end{equation}
where we use the same notation as in the main text. The probability distribution of the initial state is $p(i)=(1-e^{-\beta\,\hbar\omega})e^{-\beta\,\hbar\omega\,i}$ where $\hbar \omega$ is the single photon energy. For high average occupation number, the distribution over $i$ is smooth and we can convert the sum to an integral. We introduce the variable $x=\beta\,\hbar\omega\,i$.
 
$m=0$ corresponds to the no-photons-detected outcome and $m=1$ corresponds to the photons-detected outcome. Here we denote the probability of the $m=1$ outcome by $q$. We have for the joint probability of the initial state and measurement outcome $p(0,i)=(1-e^{-\beta\,\hbar\omega})(1-R\eta)^i e^{-\beta\,\hbar\omega\,i}$ and $p(1,i)=(1-e^{-\beta\,\hbar\omega})(1-(1-R\eta)^i) e^{-\beta\,\hbar\omega\,i}$. Converting this to a probability density in terms of the new variable $x$, we get $\mathcal{P}(0,x)=\frac{1}{\beta\,\hbar\omega}(1-e^{-\beta\,\hbar\omega})(1-R\eta)^{\frac{x}{\beta\,\hbar\omega}}e^{-x}$ and $\mathcal{P}(1,x)=\frac{1}{\beta\,\hbar\omega}(1-e^{-\beta\,\hbar\omega})(1-(1-R\eta)^{\frac{x}{\beta\,\hbar\omega}})e^{-x}$. For $\beta \ll 1$, $\frac{1}{\beta\,\hbar\omega}(1-e^{-\beta\,\hbar\omega}) \approx 1$ so we can write $\mathcal{P}(0,x)=e^{c\,x}$ and since $1-q=\int_0^\infty \mathcal{P}(0,x)\,d x=\int_0^\infty e^{c\,x}\,dx=\frac{1}{c}$, we get $c=\frac{1}{1-q}$. Finally, we have
\begin{equation}
\begin{split}
&\mathcal{P}(0,x)=e^{-\frac{1}{1-q} x}\\
&\mathcal{P}(1,x)=\left(1-e^{-\frac{q}{1-q}x}\right) e^{-x}
\end{split}
\end{equation}
and we see that these probability densities only depend on $q$. We can now compute the mutual information in the limit of high occupation numbers and high beam splitter transmittance:
\begin{equation}
\begin{split}
I=&\int\,e^{-\frac{1}{1-q} x}\left(-q/(1-q)-\log(1-q)\right)\,d x\\
+&\int\,\left(1-e^{-\frac{q}{1-q}x}\right) e^{-x}\\
&\left(\log(1-e^{-\frac{q}{1-q}x})-\log(q)\right)\,d x.
\end{split}
\end{equation}
These integrals have an analytic solution which we do not give here. The result is used to plot the bound depicted in Figure \ref{exp2}.

\end{document}